# MULTITASK BRAIN TUMOR INPAINTING WITH DIFFUSION MODELS: A METHODOLOGICAL REPORT

**Pouria Rouzrokh**[1,2,*], **Bardia Khosravi**[1,2,*],
**Shahriar Faghani**[1], **Mana Moassefi**[1], **Sanaz Vahdati**[1],
**Bradley J. Erickson**[1,+]

(1) Mayo Clinic Artificial Intelligence Laboratory, Mayo Clinic, MN, USA
(2) Orthopedic Surgery Artificial Intelligence Laboratory, Mayo Clinic, MN, USA
(*) Co-first authors, (+) Corresponding author
Please email all correspondence to: *bje@mayo.edu*

## Abstract

Despite the ever-increasing interest in applying deep learning (DL) models to medical imaging, the typical scarcity and imbalance of medical datasets can severely impact the performance of DL models. The generation of synthetic data that might be freely shared without compromising patient privacy is a well-known technique for addressing these difficulties. Inpainting algorithms are a subset of DL generative models that can alter one or more regions of an input image while matching its surrounding context and, in certain cases, non-imaging input conditions. Although the majority of inpainting techniques for medical imaging data use generative adversarial networks (GANs), the performance of these algorithms is frequently suboptimal due to their limited output variety, a problem that is already well-known for GANs. Denoising diffusion probabilistic models (DDPMs) are a recently introduced family of generative networks that can generate results of comparable quality to GANs, but with diverse outputs. In this paper, we describe a DDPM to execute multiple inpainting tasks on 2D axial slices of brain MRI with various sequences, and present proof-of-concept examples of its performance in a variety of evaluation scenarios. Our model and a public online interface to try our tool are available here.

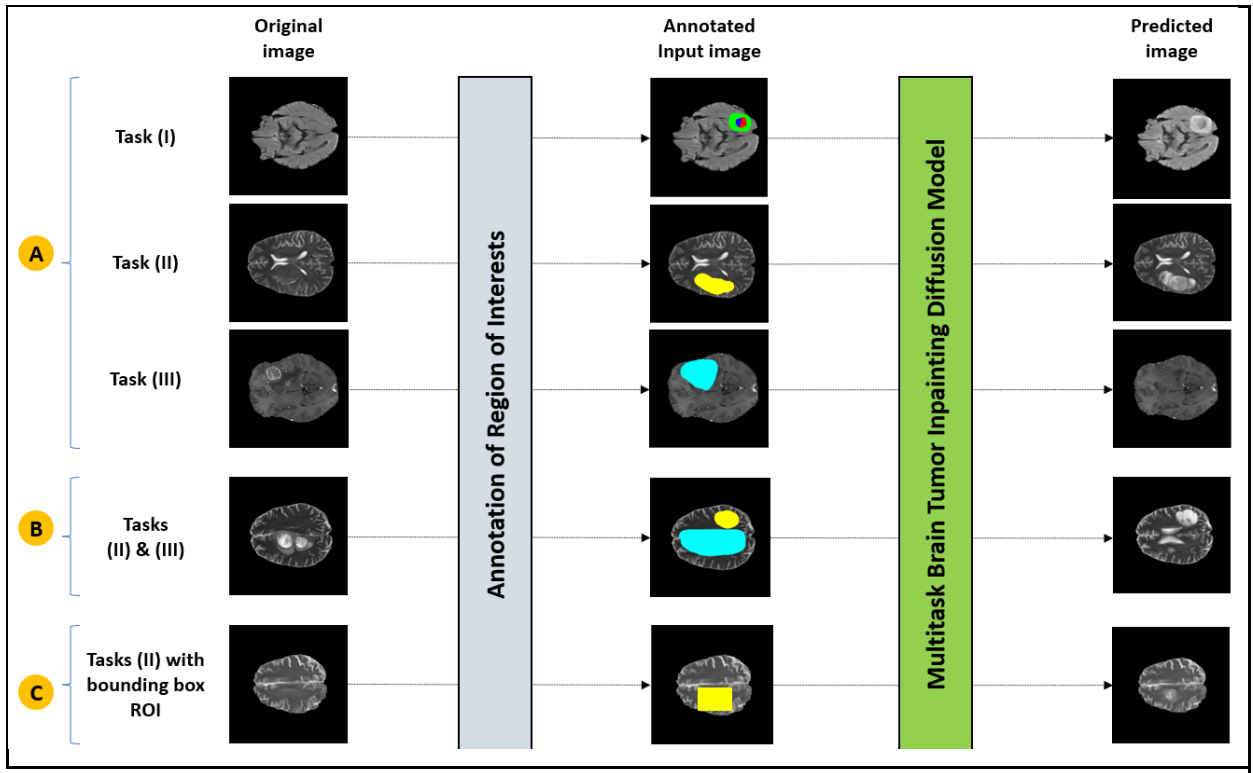

**Figure 1.** Expected target features from the multitask inpainting model. A) the model should be able to inpaint regions of interest (ROIs) for pre-determined tumoral components (task 1), a random tumor with undetermined components (task 2), or tumor-less (apparently normal) brain tissues. B) the model should be able to do tasks 1 to 3 at the same inference round. C) the model should be able to perform tasks 1 to 3 on two distinct modes of input ROIs, i.e., free-form ROIs and bounding box ROIs.

# 1 Introduction

Number of Artificial Intelligence (AI) and in particular machine learning (ML) publications related to medical imaging has expanded dramatically over the recent years(1). A recent PubMed search with the Mesh keywords "artificial intelligence" and "radiology" yielded 5,369 papers in 2021, which is more than five times the number of results from the same search in 2011. From classification to semantic segmentation, object detection, and image generation, ML models are constantly being developed to improve healthcare efficiency and outcomes(2). In diagnostic radiology, for instance, there are numerous published reports indicating that ML models may perform on par or even better than medical experts in certain tasks, such as anomaly detection and screening for pathologies(3,4). It is therefore undeniable that AI can assist radiologists and drastically cut their labor, if applied properly(5).

Despite the growing interest in developing ML models for medical imaging, there are significant challenges that can limit the practical applications of such models or even predispose them to significant bias(6–8). Two of these challenges are the issues of data scarcity and data imbalance. On the one hand, medical imaging datasets are often much smaller than the natural photograph datasets like ImageNet, and pooling institutional datasets or making them publicly available may be impossible due to patient privacy concerns. On the other hand, even those medical imaging datasets that are available to data scientists are frequently imbalanced. In other words, the volume of medical imaging data for patients with particular pathologies is substantially less than that of patients with common pathologies or healthy individuals. Training or evaluating a ML model with insufficiently large or imbalanced datasets may result in systemic biases in model performance(6).

In addition to the public release of deidentified medical imaging datasets and the endorsement of strategies such as federated learning, which facilitates machine learning (ML) model development on multi-institutional datasets without data sharing, synthetic image generation is one of the primary strategies to combat both data scarcity and data imbalance(9). Generative ML models can learn how to generate realistic medical imaging data that does not belong to a real patient and can therefore be shared publicly without compromising patient privacy. Since the emergence of generative adversarial networks (GANs), various generating models have been introduced which are capable of synthesizing high quality synthetic data(10). The majority of these models generate unlabeled imaging data that may be useful for certain use cases, such as self-supervised or semi-supervised of downstream models. Additionally, some other models are capable of conditional generation, which provides the ability to generate an image based on predetermined clinical, textual, or imaging variables. The latter group of generative models enables the production of labeled synthetic data, thereby advancing machine learning research, medical imaging quality, and patient care.

Despite the enormous success of GANs in generating synthetic medical imaging data, these models are frequently criticized for their lack of output diversity and unstable training. As a conventional alternative to GANs, autoencoder deep learning models are easier to train and able to generate more diversified outputs, but their synthetic results lack the image quality of GANs(11). Denoising Diffusion Probabilistic Models (DDPMs), or diffusion models for short, are a new class of image generation models that surpass GANs in terms of synthetic image quality and are comparable to autoencoders in terms of output diversity(12,13). Based on the Markov chain theory, diffusion models learn to generate their synthetic outputs by gradually denoising an initial image packed with random gaussian noise. This iterative denoising process makes the inference runs of diffusion models significantly slower than other generative models, but in exchange, it allows them to extract more representative features from their input data, enabling them to outperform other models in the end(14).

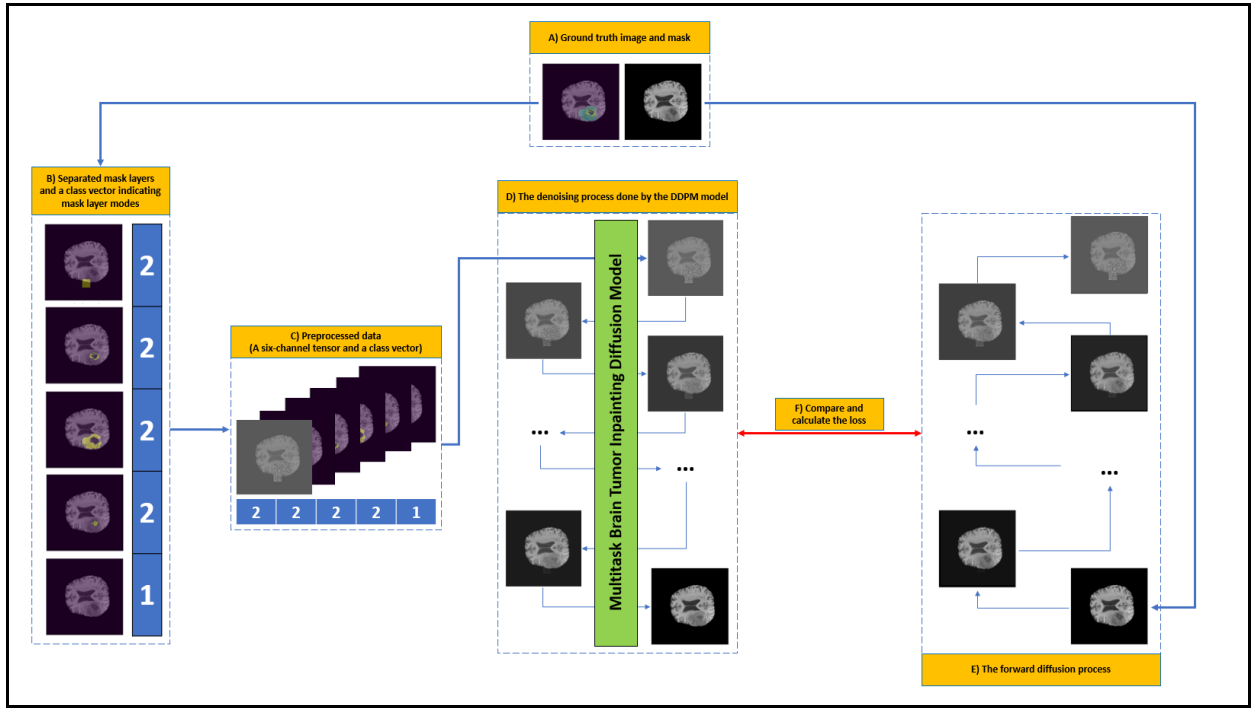

**Figure 2.** An overview of our strategy to train the multitask brain tumor inpainting algorithm. A) a pair of a ground truth image and its corresponding tumor mask is read from the training set; B) the input mask is preprocessed according to a randomization schema to generate five separate masks for distinct ROIs (from top to bottom: normal brain, necrotic tumor core, tumoral edema, tumoral enhancement, and multi-component tumor); C) the input image is preprocessed in a way that all pixels with at least one corresponding ROI are filled with random gaussian noise. This preprocessed image will be concatenated to the five mask ROIs developed in the previous step to create a six-channel tensor of size 6×256×256. Furthermore, a one-dimensional class vector is built to denote the ROI mode for each of the 5 ROI channels in the previous tensor; D) the six-channel tensor and the class vector are fed to a diffusion model to denoise the noisy image to a version that is less noisy for one step; E) The input image will similarly be converted to a ground truth noisy image at the same step that the output of the diffusion model should be; F) the output of the diffusion model will be compared with the ground truth noisy image to calculate the loss and optimize the model.

In this methodological paper, we introduce a proof-of-concept diffusion model that can be used for multitask brain tumor inpainting on multi-sequential brain magnetic resonance imaging (MRI) studies. More precisely, we developed a diffusion model that can receive a two-dimensional (2D) axial slice from a T1-weighted (T1), a contrast-enhanced T1-weighted (T1CE), a T2-Weighted (T2), or a fluid attenuated inversion recovery (FLAIR) sequence of a brain MRI and inpaint a user-defined cropped area of that slice with realistic and controllable image of either a high-grade glioma and its corresponding components (e.g., the surrounding edema), or tumor-less (apparently normal) brain tissues. The incidence of high-grade glioma is 3.56 per 100,000 population in the United States, and there are only a few publicly available MRI datasets for brain tumors(15,16). In the context of such limited data, our model will enable ML researchers to edit (induce or remove) synthetic tumoral or tumor-less tissues with configurable features on brain MRI slices.

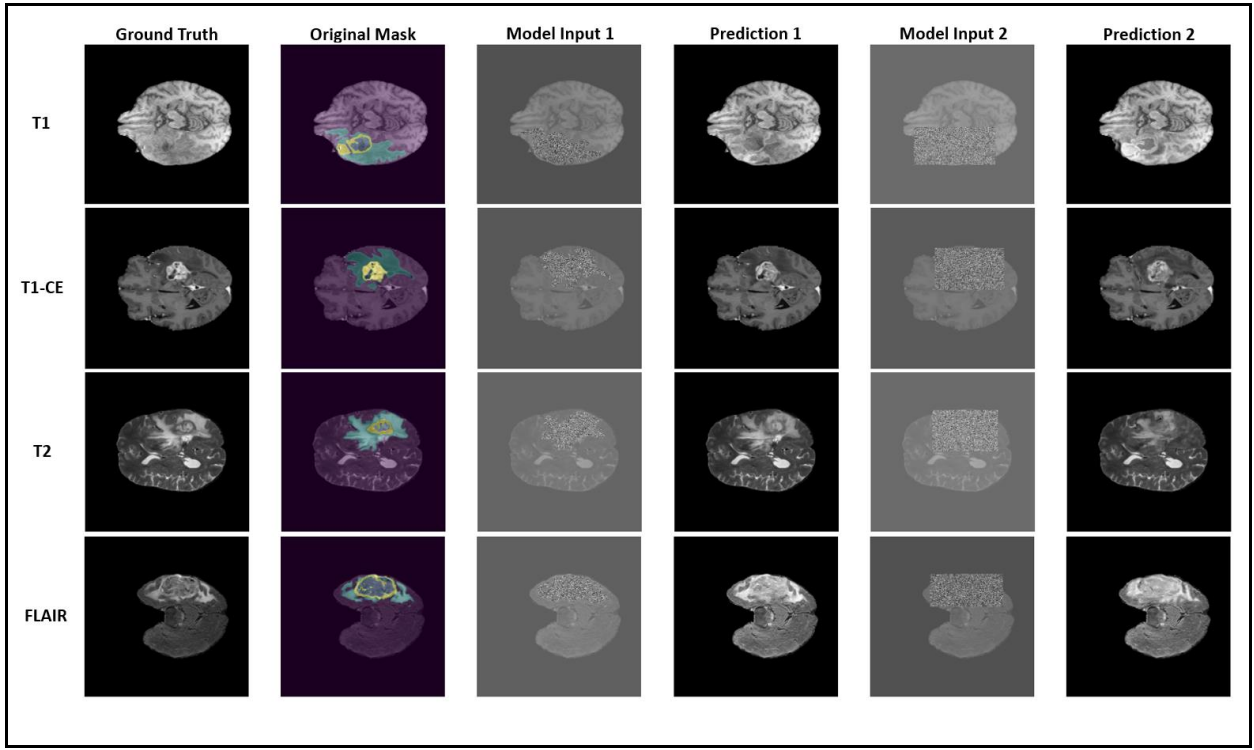

**Figure 3.** Generating tumoral lesions with predetermined regions of interest (ROIs) for necrotic tumor core, tumoral edema, and tumoral enhancement. In each instance, prediction 1 was done with a free-form input ROI and prediction 2 was done with a bounding box input ROI.

## 2 Methods

Inpainting algorithms are a group of algorithms that can remove an input image according to a user-specified mask and then fill the cropped area with synthetic imaging data that matches the surrounding context of the image. In accordance with this description, we trained a diffusion model to complete three unique sets of inpainting tasks on a 2D axial slice of a brain MRI scan with either of the T1, T1CE, T2, or FLAIR sequences: (I) to receive a mask with distinct regions of interest (ROIs) for up to three tumoral components (necrotic tumor core, tumoral edema, and tumoral enhancement) and to inpaint the input image with synthetic data matching the input ROIs for tumoral components; (II) to receive a mask with a single ROI for tumor and inpaint the input image with synthetic data matching a realistic multi-component tumoral lesion (but without predefined tumoral components); and (III) to receive a mask with a single ROI for normal brain tissue and inpaint the input image with synthetic data matching a realistic brain tissue without tumors, **Figure 1A**.

Moreover, we wanted our model to have two additional features that would increase its flexibility in downstream tasks: 1) we wanted the model to be able to perform tasks I, II, and III in the same inference round (e.g., to remove a real tumor from a brain MRI slice and to simultaneously place a synthetic tumor in the same place or elsewhere in the slice, **Figure 1B**); and 2) we wanted the model to be able to perform tasks I and II in both free-form mode and bounding-box mode. In the free-form mode, the user defines a free-form ROI for the tumoral components or the complete tumor (tasks I and II, respectively), and the model inpaints the image so that the created tumoral tissue precisely matches the ROI. In the bounding box mode, however, the user will merely define a rectangular bounding box on the input image, and the model will determine how to fill that box with tumoral tissues and tumor-free surrounding tissues. In this mode, the model does not need to fill the entire rectangle with tumor. Instead, it will inpaint a tumoral lesion and its surrounding area so that if a bounding box were fitted around the created tumoral lesion, it would match the initial bounding box specified by the user, **Figure 1C**. Lastly, we desired that the model be capable of performing inpainting in both free-form and bounding box modes simultaneously and for distinct ROIs.

Although the aforementioned objectives may sound too overwhelming to expect from a single model, we hypothesized that with the proper training setup, diffusion models are so capable of extracting relevant features and conditioning on the input data that our objectives may be achievable.

**2.1 Data**

We leveraged 1251 skull-stripped brain MRI studies from the Brain Tumor Segmentation Challenge (BraTS) which was released in 2021 by the Radiological Society of North America (RSNA), the American Society of Neuroradiology (ASNR), and the Medical Image Computing and Computer Assisted Interventions (MICCAI) society. All studies were retrieved in the Neuroimaging Informatics Technology Initiative (NIfTI) format, including a three-dimensional (3D) imaging array of shape 240×240×155, and pixel dimension of 1 mm in all three axes. Each study had at least one high-grade glioma lesion and annotated masks for different components of that lesion (necrotic tumor core, tumoral edema, and the tumoral enhancement) were provided in a separate NIfTI file corresponding to that study.

The 2D axial slices of all studies and their corresponding masks were pad-resized to 256×256 pixels and saved with the Portable Network Graphics (PNG) format. Each study had T1, T1CE, T2, and FLAIR sequences, and therefore four different versions of each axial slice were saved to disk. We then fitted a bounding box around the brain region of each slice and excluded slices whose total bounding box area was less than 100 pixels square. The remaining slices were split into training, validation, and test sets at a ratio of 8:1:1, with the size of the area occupied by tumoral lesions serving as a stratification factor. We also verified that all slices from a similar study existed in just one of the sets. This division schema resulted in 556,744 images in the training set, 70,412 images in the validation set, and 69,116 images in the test set. Finally, we oversampled the slices which had a tumoral lesion by a factor of two, increasing the size of our training set to 810,468, validation set to 102,536, and test set to 100,100 images.

During training, a pair of an axial image and its corresponding mask were read from disk and the mask was converted to a five-channel tensor of size 256×256, with channel 0 containing the ROI for normal brain tissue, channel 1 containing the ROI for necrotic tumor core, channel 2 containing the ROI for tumoral edema, channel 3 containing the ROI for tumoral enhancement, and channel 4 containing the ROI for a multi-component tumor. The construction of the five-channel tensor mirrored the separation of the original tumor mask's values into five distinct channels (with an exception of addition of the normal brain ROIs which were absent in the original tumor mask). This multichannel tensor was built following one or two preprocessing scenarios that could be selected according to a randomization scheme during the training, **Figure 2A-C**:

1. ROIs for Individual tumoral components filled channels 1 to 3 and channel 4 was left empty. Each ROI could be left with its original free-form format, or be converted to a bounding box by fitting a rectangle around the original ROI. In both instances, pixels in channels 1-3 that matched to the ROIs were assigned the value 1 and all other pixels were assigned the value 0.

2. A generic ROI was constructed by merging ROIs for all tumoral components and filling channel 4, while leaving channels 1 to 3 unfilled. Similar to the preceding case, this generic ROI might be left in its original free-form state or converted to a bounding box by fitting a rectangle around the original ROI. In both instances, channel 4 pixels that corresponded to the generic ROI were assigned a value of 1 and all other pixels were assigned a value of 0.

3. Random circular or rectangular portions of non-tumor areas were picked, and their associated masks were used to fill channel 0. Scenario 3 was compatible with scenarios 1 and 2, but was independent from them. Additionally, scenario 3 was independent of the tumor mask that could exist for a slice, as it only picked non-tumoral regions. In this case, pixels of channel 0 that corresponded to the selected ROIs were assigned the value 1, while the remaining pixels were assigned the value 0.

The preprocessing pipeline was then completed by two following steps that followed the construction of the five-channel tensor. First, all pixels of the original image that corresponded to at least one pixel with non-zero value in the multi-channel tensor were replaced with random Gaussian noise (this is done after the original image is normalized between zero to one, and then zero padded to a square shape). The image that resulted from this operation was then normalized, for a second time, to a pixel value range of -1 to 1 and concatenated with the five-channel tensor to create a six-channel tensor with dimensions 6×256×256. Second, a one-dimensional class vector of size 5 was constructed and populated with the values 1 (representing an empty channel), 2 (representing a channel filled with free-form ROI), or 3 (representing a channel filled with bounding box form ROI) based on how the five-channel tensor was constructed in the preprocessing scenarios. In other words, each class vector value represented how each channel of the five-channel mask was filled with ROIs. This class vector was utilized as a non-imaging variable to condition the diffusion model during training, and it was also meant to be used as a signal the model for the desired inpainting mode during inference (i.e., free form inpainting or bounding box inpainting).

The six-channel tensor and the one-dimensional class-vector were the end outcomes of preprocessing the initial brain slice image and its accompanying tumor mask. The data loader created batches of these two tensors, which were then supplied to the diffusion model during training.

## 2.2 Model

We trained a diffusion model capable of transforming an input image of an axial brain MRI slice with one or more areas of random noise (channel 0 of the six-channel tensor above) into a synthetic yet realistic image with the user-specified attributes. This conversion occurs across 1000 steps of denoising and is conditioned on the input ROIs to the model (channels 1 through 5 of the six-channel tensor) and the one-dimensional class vector that specifies the mode of each ROI.

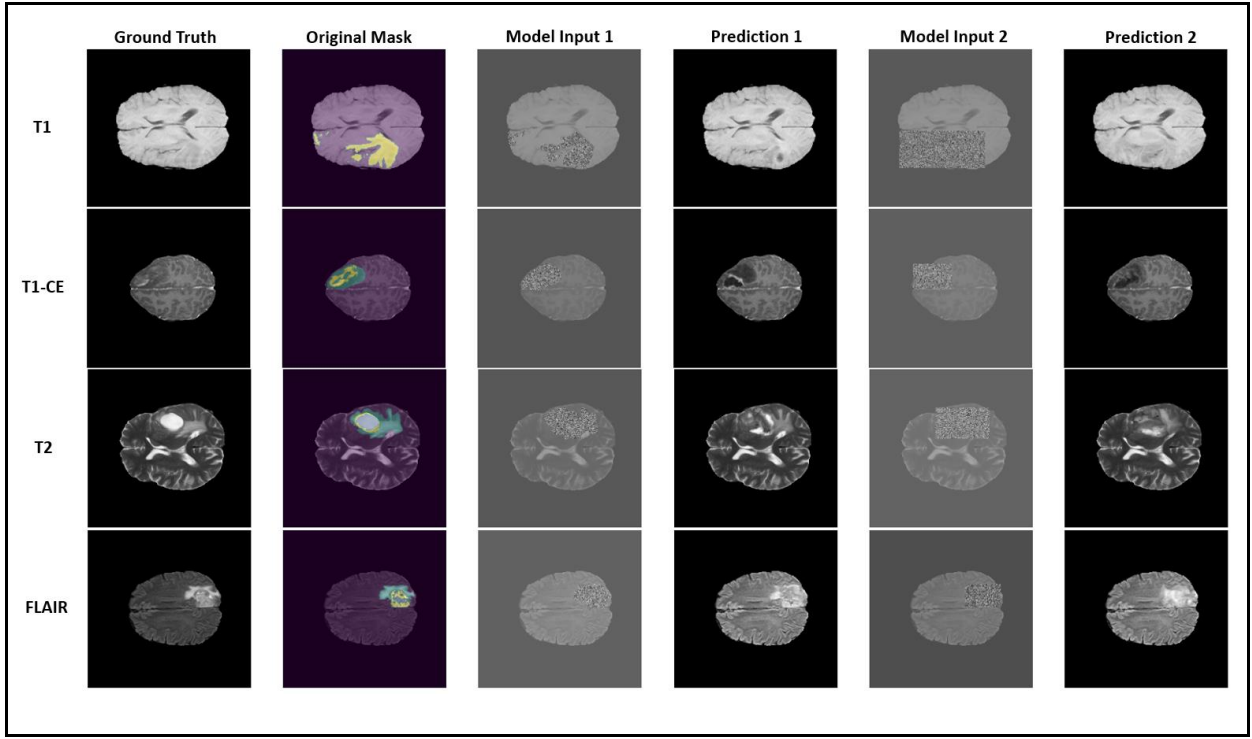

**Figure 4.** Generating tumoral lesions with undetermined (random) regions of interest (ROIs) for necrotic tumor core, tumoral edema, and tumoral enhancement. In each instance, prediction 1 was done with a free-form input ROI and prediction 2 was done with a bounding box input ROI.

While the mathematical details of diffusion models are discussed elsewhere(12,13,17), a high-level summary of how diffusion models work is as follows. In general, one can follow the Markov chain rule to add random gaussian noise to an input image for a finite number of time ($T$ steps) until all the image's signal is destroyed and the image is transformed into random Gaussian noise of the same size as the original image. This process is referred to as forward diffusion and can be done without using a deep learning model. In other words, the noisy image of step n, where $n \in [1, T]$, can be mathematically obtained with simple coding. However, the reverse transitions of this chain could be learned by a deep learning model with a UNet-like architecture. In this way, the model will act as a parametrized reverse Markov chain to receive a noisy image of step *n*, where $n \in [1, T]$, and generates a less noisy version of that image at step n-1. During training of such a model, a noisy version of the training data is generated at random steps, and the model learns to denoise each noisy data point for one step.

We implemented our inpainting model aligned with the above high-level definition, but with several adjustments, **Figure 2D-F**:
9

First, the loss of our model was not calculated based on denoising the entire image, but only an area of the image that was originally filled with random gaussian noise (per user annotation). As a result, our diffusion model was trained to be an inpainting algorithm that could rely on the surrounding context of a noisy area for denoising that.

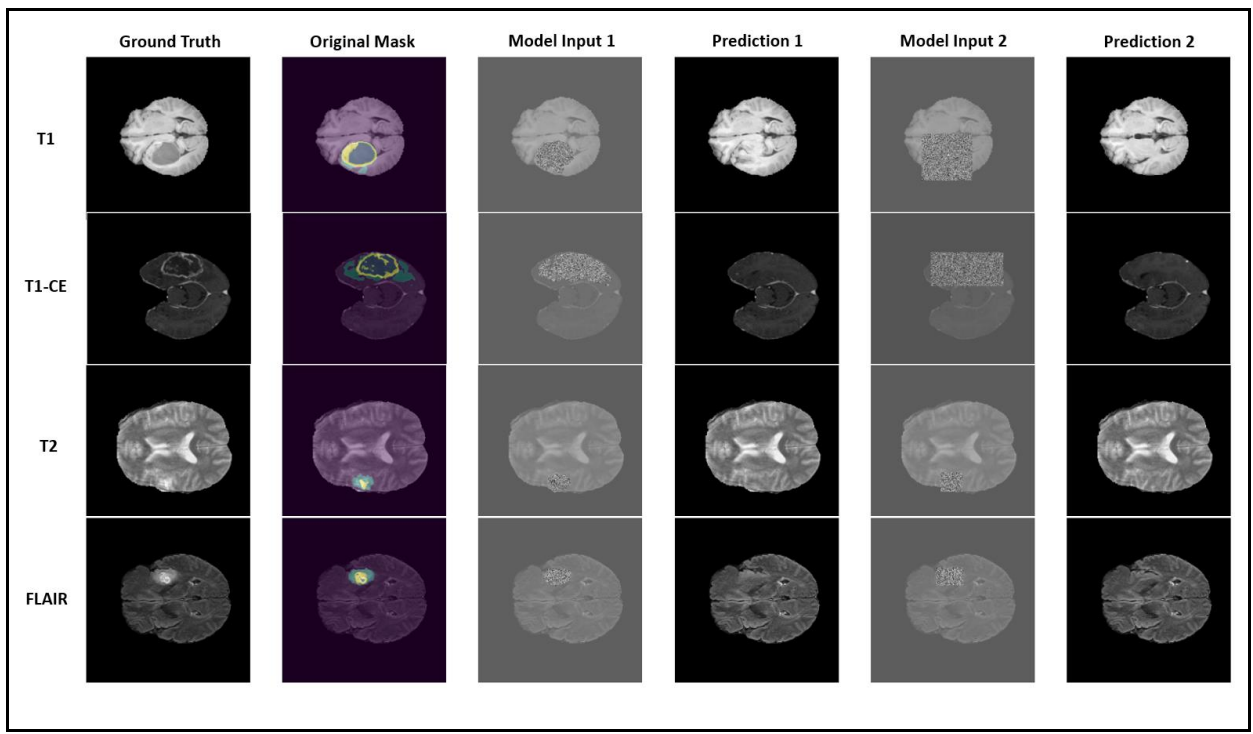

**Figure 5.** Generating tumor-less (apparently normal) brain tissues instead of the regions of interest (ROIs) for tumoral lesions. In each instance, prediction 1 was done with a free-form input ROI and prediction 2 was done with a bounding box input ROI.

Second, we adopted a UNet-like algorithm originally introduced by Dhariwal et al.(17), but modified it to receive six additional channels (the noisy image and five layers of ROIs) for the input image, and output a single channel tensor (the inpainted image). Moreover, we slightly modified the conditioning mechanism of the original UNet-like algorithm on the one-dimensional class vector, so that the embeddings of that vector were concatenated with the timestep embeddings of the original UNet-like algorithm, not added to it. The source code for the modified UNet-like model is available in our GitHub repository[1].

Third, we followed the classifier-free guidance protocol, originally proposed by Ho et al.(18), to condition our model on the one-dimensional class vector. Per this protocol, all values in the one-dimensional vector were randomly imputed with zero values in 10% of the training data. While this modification did not change how our diffusion model was trained, it enabled us to adjust the model's conditioning on user ROIs during the inference. The adjustment was possible by using the following formula during the inference, where $O$ is the final model output, $P_{Initial}$ is the model output with a user-determined class vector, and $P_0$ is the model output for the similar input data but a class vector with zero values in all elements. $W$ is a positive integer value denoting the conditioning weight.

---

[1] https://github.com/Mayo-Radiology-Informatics-Lab/MBTI

$$O = (W+1)P_{Initial} - P_0$$

Increasing the conditioning weight during inference will compel our model to generate synthetic images that place a greater focus on the user-input ROIs, similar to a truncation procedure that is occasionally used in post-processing of GAN outputs.

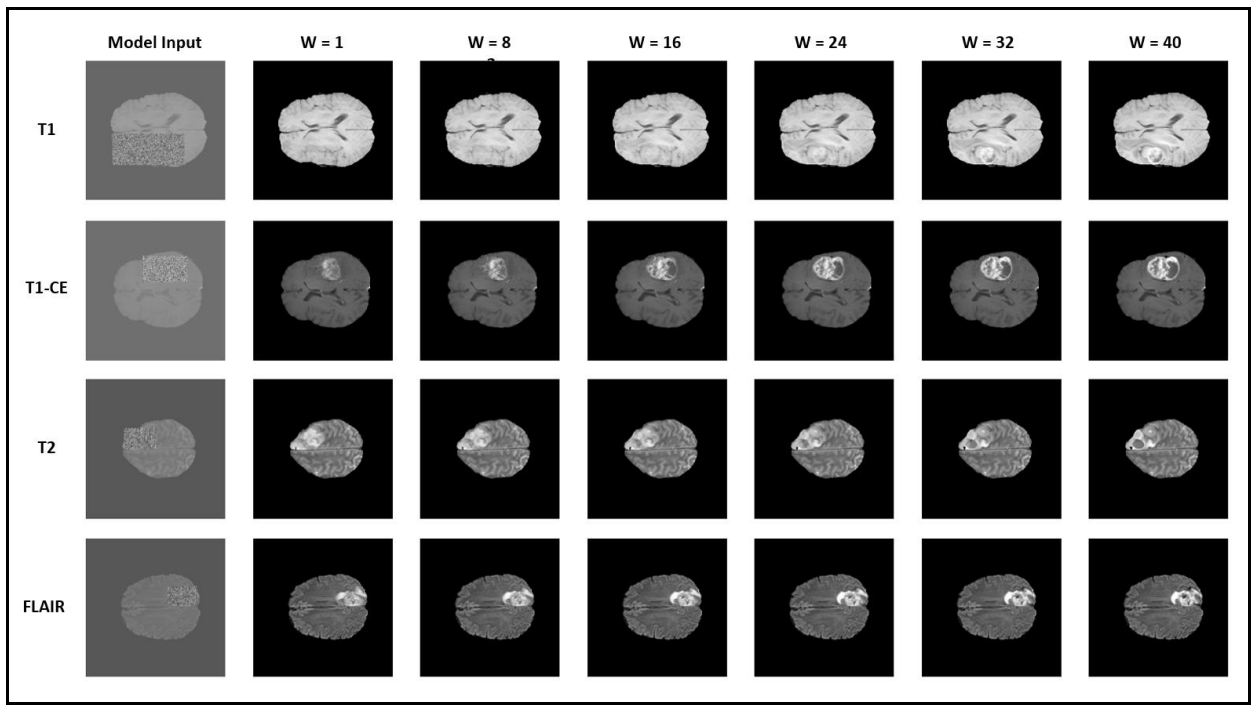

**Figure 6.** Multiple attempts to generate tumoral lesions with undetermined (random) regions of interest (ROIs) for necrotic tumor core, tumoral edema, and tumoral enhancement. We increase the conditioning weight for the model's prediction while keeping the same random seed across different attempts.

Finally, we reconfigured our trained diffusion model as a Denoising Diffusion Implicit Model (DDIM) for inference rounds, following a protocol originally introduced by Song et. al(19). In contrast to DDPM models, which must denoise their input noise using the exact same number of denoising steps with which they were trained, a DDIM model can be modified to complete the inference round in fewer denoising steps, at the expense of a minimal drop in image quality.

**2.3 Training**

Our model was trained with a batch size of 108 using three A100 NVIDIA Graphic Processing Unit cards, each with 80 gigabytes of random-access memory, and a distributed data parallel protocol. We continued the training with a constant learning rate of 1e-4 until the model on 40 million images (which is equal to 370,000 iterations). An exponential moving average instance of model weights was updated at each iteration during the training and was ultimately used for inference. We used random axial and horizontal flipping, and random rotations in the range of -15 to 15 degrees as the only augmentations during training. The model, training, and inference pipeline were all coded in PyTorch v1.11.

# 3 Performance Evaluation

For this methodological report, we only provide visual examples of our model's performance and leave out the quantitative evaluations for an upcoming separate report. We will demonstrate our model's performance in five different scenarios, while images used in all scenarios are randomly selected from the test set, and the four rows of figures corresponding to each scenario represent slices with T1, T1CE, T2, and FLAIR weightings, in descending order. Unless otherwise noted, the conditioning weight was set to 0.4 for all scenarios and the predictions were made by 1000 steps of denoising. A public online interface to try our tool is available at our GitHub page[2].

*Scenario 1*, **Figure 3**: In this scenario, our model was utilized to generate a synthetic lesion with similar components to the true lesion in the input slice. To do this, the available mask for each slice was preprocessed and supplied to the model with distinct ROIs for various tumoral components. Two distinct modes of prediction are illustrated. In the first mode, ROIs were input to the model in free-form format (i.e., precisely the same as the original ROIs in the BraTS dataset), and it was anticipated that the model would generate a tumor that was comparable to the original lesion. In the second mode, the ROIs for necrotic tumor core and tumoral enhancement were preprocessed similarly to the first mode, but the ROI for tumoral edema was changed to a bounding box that encompassed the original ROI for that component. This second mode, as anticipated, forced our model to create random ROIs for the tumoral edema that nonetheless fit within the input bounding box.

---

[2] https://github.com/Mayo-Radiology-Informatics-Lab/MBTI

*Scenario 2*, **Figure 4**: This experiment was similar to scenario 1, with one exception: the ROIs for all tumoral components were combined into a single ROI, and the model was instructed to inpaint a random multi-component tumor on the input slice, with no restrictions on the exact tumoral component borders. Similar to scenario 1, the first mode of prediction required the model to inpaint a tumor in a free-form manner, but the second mode of prediction required the model to inpaint a tumor and its surrounding tissue within a bounding box.

*Scenario 3*, **Figure 5**: This experiment was similar to Scenario 2, but this time the model was instructed to inpaint normal brain tissue as opposed to tumorous lesions. Due to the fact that the initial tumor could have had a compressive effect on its surrounding tissue that was not annotated in the tumor mask, the bounding box mode of prediction produced more realistic results than the free-form mode in this scenario. This is explained by the fact that a bounding box ROI will encompass more of the input slice and the model can reproduce the original tumor's surrounding area in addition to the tumor itself.

*Scenario 4*, **Figure 6**: To visualize the influence of the conditioning weight on our model's outputs, we repeated scenario 2 (using the bounding box mode for ROIs) six times with a constant randomization seed, but with increasing the conditioning weight from w=1 in prediction 1 to w=40 in prediction 6. As previously mentioned, increasing the conditioning weight caused the model to place a greater focus on tumor signals in the generated images (e.g., by modifying the intensity of created ROIs), but could also add watercolor artifacts and make the outputs appear unrealistic in higher weights.

*Scenario 5*, **Figure 7**: Similar to variational autoencoder models, a diffusion model's outputs can vary by altering the randomization seed. In this scenario, the experiment from scenario 2 (using the bounding box mode for ROIs) is repeated, but with six different randomization seeds. As depicted in the figure, the model's outputs are considerably different across the six predictions, indicating the potential of our inpainting model to provide diverse outputs.

## 4  Discussion

We presented a methodology and proof-of-concept findings for leveraging diffusion models to construct a multitask brain tumor inpainting algorithm. Our model is capable of performing multiple inpainting jobs (creating individual tumoral components, a multi-component tumor, or normal-appearing brain tissue) in a single inference round. It can also perform tumor inpainting on two distinct types of input ROIs, the free form ROI and the bounding box ROI, where for the former type it will inpaint a lesion exactly matching the borders of the input ROI, and for the latter type it will generate a random lesion and its surrounding tissue so that they fit together within the input bounding box. In addition, we demonstrated multiple capabilities of our model, such as its ability to generate infinite instances of synthetic images for a unique input with different randomization seeds, to alter the emphasis on user-defined ROIs in the generated image by adjusting the conditioning weight, and to have short inference runs by endorsing a DDIM protocol for inference.

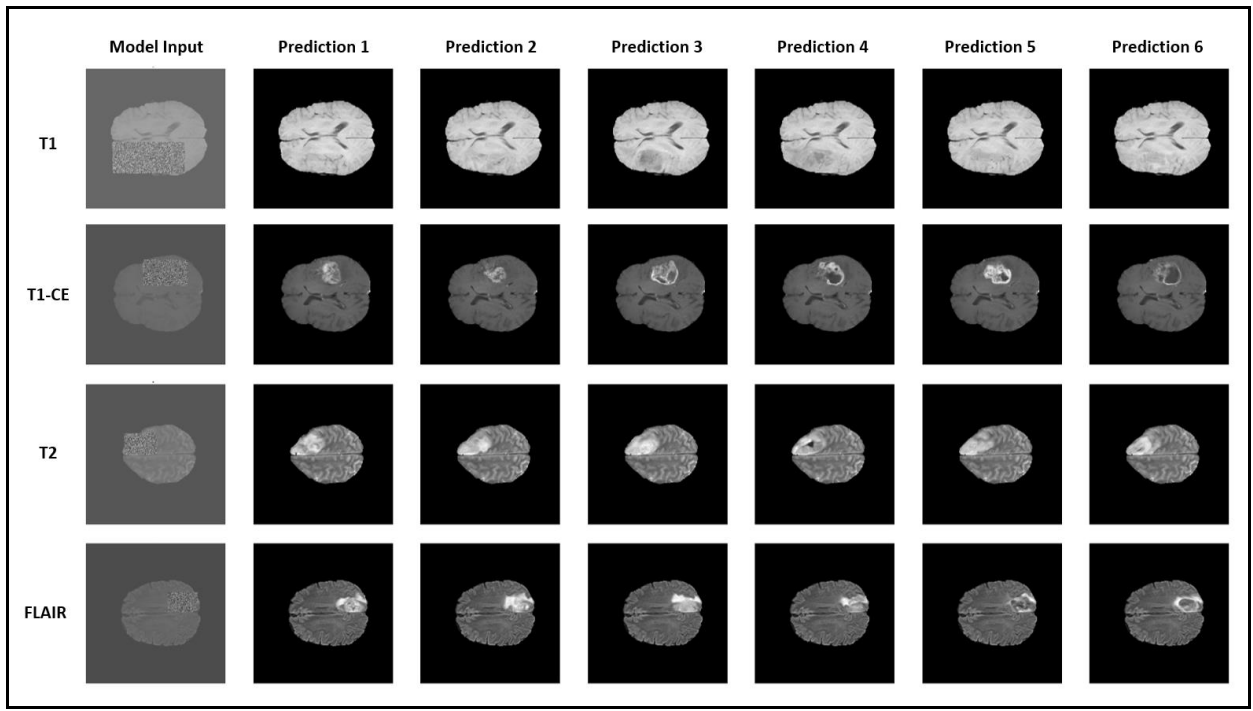

**Figure 7.** Multiple attempts to generate tumoral lesions with undetermined (random) regions of interest (ROIs) for necrotic tumor core, tumoral edema, and tumoral enhancement. We change the random seed across different attempts to produce diverse results.

Inpainting algorithms are a family of deep learning models that are generally employed for image editing tasks such as the removal of artifacts in natural pictures(20). Although a number of studies have presented deep learning methods for inpainting medical imaging data(21–24), the vast majority of these algorithms employ GANs as their fundamental generative models. However, GANs frequently provide outputs with limited diversity, which could negatively impact the performance of inpainting algorithms. Using diffusion models as inpainting methods will address this important obstacle and open the door to the production of high-quality, diverse synthetic images. Despite the fact that a published example of employing diffusion models for inpainting brain tumors on MRI data existed at the time of the publication of this paper(25), our multitask inpainting approach offers far more flexibility than the earlier algorithm. Our multitask algorithm can be used to construct realistic synthetic brain MRIs with configurable tumoral and non-tumoral tissues, allowing researchers to develop and share databases of synthetic data without compromising patient confidentiality. Last but not least, our method might be used to build datasets with engineered properties of lesions, giving a benchmark for evaluating the capabilities of deep learning models trained on our dataset, such as their calibration and uncertainty quantification.

In summary, we developed a multitask brain tumor inpainting algorithm capable of inpainting tumoral lesions and tumor-free brain tissue in 2D slices of brain MRI studies with T1, T1CE, T2, or FLAIR sequences. We showed proof-of-concept examples of our model's performance that suggest its potential for creating synthetic, yet realistic, brain tumor MRI datasets for different use cases.

# 5 References

1. Barragán-Montero A, Javaid U, Valdés G, et al. Artificial intelligence and machine learning for medical imaging: A technology review. Phys Med. 2021;83:242–256.

2. van Leeuwen KG, de Rooij M, Schalekamp S, van Ginneken B, Rutten MJCM. How does artificial intelligence in radiology improve efficiency and health outcomes? Pediatr Radiol. 2022;52(11):2087–2093.

3. Shiraishi J, Li Q, Appelbaum D, Doi K. Computer-aided diagnosis and artificial intelligence in clinical imaging. Semin Nucl Med. 2011;41(6):449–462.

4. Leibig C, Brehmer M, Bunk S, Byng D, Pinker K, Umutlu L. Combining the strengths of radiologists and AI for breast cancer screening: a retrospective analysis. Lancet Digit Health. 2022;4(7):e507–e519.

5. Sorantin E, Grasser MG, Hemmelmayr A, et al. The augmented radiologist: artificial intelligence in the practice of radiology. Pediatr Radiol. 2022;52(11):2074–2086.

6. Rouzrokh P, Khosravi B, Faghani S, et al. Mitigating Bias in Radiology Machine Learning: 1. Data Handling. Radiology: Artificial Intelligence. Radiological Society of North America; 2022;4(5):e210290.

7. Zhang K, Khosravi B, Vahdati S, et al. Mitigating Bias in Radiology Machine Learning: 2. Model Development. Radiology: Artificial Intelligence. Radiological Society of North America; 2022;e220010.

8. Faghani S, Khosravi B, Zhang K, et al. Mitigating Bias in Radiology Machine Learning: 3. Performance Metrics. Radiology: Artificial Intelligence. Radiological Society of North America; 2022;e220061.

9. Yu B, Wang Y, Wang L, Shen D, Zhou L. Medical Image Synthesis via Deep Learning. Adv Exp

Med Biol. 2020;1213:23–44.

10. Wang L, Chen W, Yang W, Bi F, Yu FR. A State-of-the-Art Review on Image Synthesis With Generative Adversarial Networks. IEEE Access. 2020;8:63514–63537.

11. Bond-Taylor S, Leach A, Long Y, Willcocks CG. Deep Generative Modelling: A Comparative Review of VAEs, GANs, Normalizing Flows, Energy-Based and Autoregressive Models. IEEE Trans Pattern Anal Mach Intell. 2022;44(11):7327–7347.

12. Nichol A, Dhariwal P. Improved denoising diffusion probabilistic models. arXiv [cs.LG]. 2021. http://arxiv.org/abs/2102.09672.

13. Ho J, Jain A, Abbeel P. Denoising Diffusion Probabilistic Models. arXiv [cs.LG]. 2020. http://arxiv.org/abs/2006.11239.

14. Baranchuk D, Rubachev I, Voynov A, Khrulkov V, Babenko A. Label-Efficient Semantic Segmentation with Diffusion Models. arXiv [cs.CV]. 2021. http://arxiv.org/abs/2112.03126.

15. Baid U, Ghodasara S, Mohan S, et al. The RSNA-ASNR-MICCAI BraTS 2021 Benchmark on Brain Tumor Segmentation and Radiogenomic Classification. arXiv [cs.CV]. 2021. http://arxiv.org/abs/2107.02314.

16. Calabrese E, Villanueva-Meyer JE, Rudie JD, et al. The University of California San Francisco Preoperative Diffuse Glioma MRI Dataset. Radiology: Artificial Intelligence. Radiological Society of North America; 2022;4(6):e220058.

17. Dhariwal P, Nichol A. Diffusion models beat GANs on image synthesis. arXiv [cs.LG]. 2021. p. 8780–8794. https://proceedings.neurips.cc/paper/2021/hash/49ad23d1ec9fa4bd8d77d02681df5cfa-Abstract.html. Accessed October 20, 2022.

18. Ho J, Salimans T. Classifier-Free Diffusion Guidance. arXiv [cs.LG]. 2022. http://arxiv.org/abs/2207.12598.

19. Song J, Meng C, Ermon S. Denoising Diffusion Implicit Models. arXiv [cs.LG]. 2020. http://arxiv.org/abs/2010.02502.

20. Qin Z, Zeng Q, Zong Y, Xu F. Image inpainting based on deep learning: A review. Displays. 2021;69:102028.

21. Manjón JV, Romero JE, Vivo-Hernando R, et al. Blind MRI Brain Lesion Inpainting Using Deep Learning. Simulation and Synthesis in Medical Imaging. Springer International Publishing; 2020. p. 41–49.

22. Armanious K, Mecky Y, Gatidis S, Yang B. Adversarial Inpainting of Medical Image Modalities. ICASSP 2019 - 2019 IEEE International Conference on Acoustics, Speech and Signal Processing (ICASSP). 2019. p. 3267–3271.

23. Armanious K, Kumar V, Abdulatif S, Hepp T, Gatidis S, Yang B. ipA-MedGAN: Inpainting of Arbitrary Regions in Medical Imaging. 2020 IEEE International Conference on Image Processing (ICIP). 2020. p. 3005–3009.

24. Peng C, Li B, Li M, et al. An irregular metal trace inpainting network for x-ray CT metal artifact reduction. Med Phys. 2020;47(9):4087–4100.

25. Wolleb J, Sandkühler R, Bieder F, Cattin PC. The Swiss Army Knife for Image-to-Image Translation: Multi-Task Diffusion Models. arXiv [cs.CV]. 2022. http://arxiv.org/abs/2204.02641.